\documentclass[%
 reprint,
 superscriptaddress,
 selectp,
 amsmath,amssymb,
 aps,
 prapplied,
]{revtex4-2}
\usepackage{selectp}
\usepackage{mathrsfs}

\usepackage{chngcntr}
\usepackage{hyperref}
\usepackage{siunitx}
\usepackage{graphicx}
\usepackage{dcolumn}
\usepackage{bm}

\begin{document}

\title{Sign reversal of the effective Hall coefficient in laminates}

\author{Christian Kern}
\email{physics@chrkern.de}
\affiliation{Samson \& Partner Patentanwälte mbB, Widenmayerstraße 6, 80538 Munich, Germany}

\begin{abstract}
In the theory of composites, hierarchical laminate microstructures are known to often show optimal behavior. In this paper, their performance in the context of the Hall effect is evaluated. Using numerical calculations, it is shown that---despite the fact that they are the result of a simple layering process---they can mimic the behavior of chain-mail-inspired composites, which exhibit a sign-inversion of the effective Hall coefficient. To obtain such a hierarchical laminate, a two-step strategy is used: In the first step, a rank-three laminate with an effective $S$-tensor that has a sign-inverted trace is introduced. In the second step, the final isotropic hierarchical laminate is obtained from the rank-three laminate using an idea of Schulgasser's. As measured by the conductivity contrast required for the inversion as well as by the modulus of the obtained sign-inverted Hall coefficient, the identified hierarchical laminate performs better than the previously studied chain-mail-inspired composites.
\end{abstract}

\maketitle

\section{Introduction}
Composite materials are microscopically structured materials that effectively act like homogeneous materials on the macroscopic length scale \cite{Milton02}. One of the most interesting aspects of composites is the fact that their effective properties can be very different from the properties of their constituent materials, i.e., the materials from which they are made at the microscopic length scale \cite{Milton02, Kadic_2013, Kadic19}. Particularly interesting are instances in which an effective material parameter and the corresponding material parameters of the constituent materials have opposite sign. Corresponding examples cover a wide range of fields \cite{Lakes87, Milton:1992:CMP, Lakes96, Mei06, Lakes08, Gatt08, Yang08, Nicolaou12, Kadic_2013, Qu17-1, Qu17-2}. Here, with the sign-inversion of the effective Hall coefficient, an example concerning electric conduction in the presence of a magnetic field is considered.

The Hall effect describes, in its simplest form, the occurrence of a transversal voltage, the so-called Hall voltage, in a current-carrying slab-like conductor that is subject to a magnetic field \cite{Hall1879, Popovic03}. Mathematically, it is a manifestation of an antisymmetric contribution to the conductivity tensor that is the result of the magnetic field locally breaking time-reversal symmetry, implying that it is a nonreciprocal effect \cite{Onsager31, LandauStat, LandauCont, Caloz18}. Aside from being employed in many sensing applications \cite{Heremans_1993, Ripka01}, the Hall effect has been used for many decades to study the properties of semiconductors, see, e.g., Chap.\,6 in Ref.\,\onlinecite{Popovic03} for an introduction. For example, the sign of the Hall coefficient is usually determined by the predominant type of charge carriers. Thus, from the sign of the measured Hall voltage, which in a conventional Hall bar is given by the sign of the Hall coefficient, one can tell, whether the conduction in a given sample is due to electrons or holes.  

As the mathematically rigorous construction of a composite exhibiting a sign-inversion of the effective Hall coefficient by Briane and Milton \cite{Briane09}, which was based on earlier theoretical work \cite{Briane:2004:CSC}, as well as experimental realizations \cite{Kern:2016:EES, Kern19} of a simplified design \cite{Kadic:2015:HES} have shown, this simple textbook link between the Hall coefficient and the underlying conduction type does not hold true for composites. The key to obtaining the sign-inversion was a rather intricate geometry based on a periodic arrangement of intertwined tori that was inspired by medieval chain mail. Later, a second composite with a complicated topology leading to a sign-inversion of the effective Hall coefficient was identified \cite{Kern18-1}. 

This paper is concerned with the Hall effect in laminates, which form a particularly simple yet very powerful class of composites. A detailed introduction to laminates is given in Chap.\,9 of Ref.\,\onlinecite{Milton02}. The simplest type of laminate, a so-called rank-one laminate, is formed by periodically layering two or more media in a given direction. One can apply this procedure iteratively, i.e., each of the media might be a laminate itself, thereby obtaining a laminate of higher rank (a so-called hierarchical laminate), a concept dating back to Maxwell \cite{Maxwell:1954:TEM}. Laminates have the advantage over most other composites that their effective properties can be readily calculated, which facilitates a mathematically rigorous treatment. Furthermore, despite their simple building principle, they attain many fundamental bounds, i.e., they often realize the most extreme effective properties any composite can fundamentally have \cite{Milton86, Briane04, Kern20}. 

The effective properties of laminates extend in fact so far that it is difficult to identify a scenario in which their range (the so-called lamination closure, $G_{\text{L}}$) is smaller than the range of effective properties of more general composites (the so-called $G$-closure) and the border between those two regimes is yet to be better understood \cite{Milton20}. This problem is also of interest from a more mathematical perspective, as it is connected (see Chap.\,31 in Ref.\,\onlinecite{Milton02}) to the relation between rank-one convexity and quasiconvexity, which plays an important role in the calculus of variations \cite{Dacorogna07}. In Chap.\,31.9 in Ref.\,\onlinecite{Milton02}, based on \v{S}ver\'ak's example of a function that is rank-one convex but not quasiconvex \cite{Sverak92}, Milton has introduced a three-dimensional composite made from seven constituent materials with fixed orientation that has effective elastic properties that cannot be attained by a laminate. However, it remains to be answered, whether there are simpler instances of the properties of general composites going beyond those of laminates, especially with a smaller number of constituent materials and without keeping their orientation fixed \cite{Milton20}.

The sign-inversion of the effective Hall coefficient seems to be a natural candidate. Given the complicated topology of the chain-mail-inspired microstructure and the fact that its design was based on the determinant of the matrix-valued microscopic electric field (see below for a definition) having locally negative values \cite{Briane09}, an effect that cannot occur in laminates \cite{Briane04}, one would expect that the effective Hall coefficient cannot be sign-inverted in isotropic laminates, at least not in such made from two constituent materials~\cite{Milton20}. 

In the following, after a brief introduction to the fundamentals and some considerations concerning individual components of the effective Hall tensor, it is shown that, contrary to this expectation, there are isotropic hierarchical laminates made from two isotropic constituent materials exhibiting a sign-inversion of the effective Hall coefficient. Moreover, it is shown that the identified hierarchical laminate allows for a more strongly pronounced sign-inversion than the chain-mail-inspired composite. 

\section{Fundamentals}
The first part of this section forms a brief introduction to the theoretical description of the Hall effect in composites. Details can be found in Refs.\,\onlinecite{Milton02, Briane09}. 
It is then described, how the effective properties of the hierarchical laminates presented here were calculated. Finally, the necessity of increasing the complexity of the hierarchical laminates beyond the usually sufficient degree is briefly discussed. 

\subsection{The Hall effect in composites}
We consider the electric conductivity problem,
\begin{equation}
\nabla\cdot\tilde{\bm{j}}=0, ~\nabla\times\tilde{\bm{e}}=0, ~\tilde{\bm{j}}=\tilde{\bm{\sigma}}(\bm{b})\tilde{\bm{e}},
\end{equation}
in the presence of a magnetic field $\bm{b}$. Here, $\tilde{\bm{j}}$ is the electric current density, $\tilde{\bm{e}}$ is the electric field, and $\tilde{\bm{\sigma}}(\bm{b})$ is the magnetic-field dependent conductivity tensor. Throughout this paper, it is assumed that the magnetic field is weak, in which case the most general form of the constitutive law is given by, see Refs.\,\onlinecite{LandauCont, Briane09},
\begin{equation}
\tilde{\bm{j}}=\bm{\sigma}\tilde{\bm{e}}+(\bm{S}\bm{b})\times\tilde{\bm{e}}
\label{eq:constrel}
\end{equation}
and its inverted form (keeping terms up to the first order in the magnetic field) reads
\begin{equation}
\tilde{\bm{e}}=\bm{\rho}\tilde{\bm{j}}+(\bm{A}\bm{b})\times\tilde{\bm{j}},
\end{equation}
where $\bm{\sigma}$ and $\bm{\rho}$ are the zero magnetic-field (zmf) conductivity and zmf\ resistivity tensor, respectively, and $\bm{S}$ and $\bm{A}$ are rank-two tensors, which are being referred to as the $S$-tensor and the Hall-tensor, respectively, and which are related via (see Prop.\,3 in Ref.\,\onlinecite{Briane09})
\begin{equation}
\bm{S} = -\text{Cof}\left(\bm{\sigma}\right)\bm{A},
\label{eq:cof}
\end{equation}
where $\text{Cof}(\cdot)$ denotes the cofactor matrix. For an isotropic material, the tensors $\bm{A}$ and $\bm{S}$ reduce to scalar multiples of the identity. Specifically, the Hall tensor of an isotropic material is given by $\bm{A}=A\bm{I}$, where $A$ is the Hall coefficient.\\

We are interested in the effective zmf\ conductivity tensor, $\bm{\sigma}^*$, and effective $S$-tensor, $\bm{S}^*$, or, equivalently, the effective zmf\ resistivity tensor, $\bm{\rho}^*$, and effective Hall-tensor, $\bm{A}^*$ of the composite, which are defined via the macroscopic version of the constitutive law and its inverse version, 
\begin{align}
\langle\tilde{\bm{j}}\rangle&=\bm{\sigma}^*\langle\tilde{\bm{e}}\rangle+(\bm{S}^*\bm{b})\times\langle\tilde{\bm{e}}\rangle\\
\text{and }\langle\tilde{\bm{e}}\rangle&=\bm{\rho}^*\langle\tilde{\bm{j}}\rangle+(\bm{A}^*\bm{b})\times\langle\tilde{\bm{j}}\rangle,
\end{align}
respectively, where the macroscopic fields $\langle\tilde{\bm{e}}\rangle$ and $\langle\tilde{\bm{j}}\rangle$ are obtained from the corresponding microscopic fields by averaging over the unit cell of periodicity. Analogously to Eq.\,(\ref{eq:cof}), the effective $S$-tensor and the effective Hall tensor are linked via 
\begin{equation}
\bm{S}^* = -\text{Cof}\left(\bm{\sigma}^*\right)\bm{A}^*.
\label{eq:cofeff}
\end{equation}\\[-0.6cm]

The effective zmf\ conductivity tensor can be determined by evaluating the macroscopic constitutive law
\begin{equation}
\langle\bm{j}\rangle=\bm{\sigma}^*\langle\bm{e}\rangle
\label{eq:constrel_mac}
\end{equation}
for (in three dimensions) three different pairs of fields $\bm{e}=\bm{e}_{(i)}$, $\bm{j}=\bm{\sigma}\bm{e}_{(i)}$ with $i\in\lbrace 1,2,3\rbrace$ solving the conductivity problem in the absence of a magnetic field (the absence is indicated by omitting the tilde). It is convenient to assume that the three solutions for the electric field satisfy $\langle\bm{e}_{(i)}\rangle=\hat{\bm{x}}_i$, i.e., that the macroscopic fields are given by the three unit vectors, and to introduce a matrix-valued field $\bm{E}$, whose columns are given by the fields $\bm{e}_{(i)}$. The field $\bm{E}$ is the so-called matrix-valued electric field (in homogenization theory it is commonly termed corrector matrix \cite{Briane:2004:CSC, Murat:1997:C}) and gives the microscopic electric field in the composite for any choice of the macroscopic electric field,
\begin{equation}
\bm{e}=\bm{E}\langle\bm{e}\rangle.
\end{equation}
In terms of the matrix-valued electric field, the effective zmf\ conductivity tensor is given by
\begin{equation}
\bm{\sigma}^*=\langle\bm{\sigma}\bm{E}\rangle.
\end{equation}

Assuming that the magnetic field is weak, one can obtain not only the zmf\ conductivity tensor, but also the effective $S$-tensor by solving the conductivity problem for zero magnetic field. More precisely, using a perturbation approach \cite{Briane09}, see also Chap.\,16 in Ref.\,\onlinecite{Milton02}, one obtains the following expression for the effective $S$-tensor, see Thm.\,3 in Ref.\,\onlinecite{Briane09},
\begin{equation}
\bm{S}^{*}=\langle\text{Cof}(\bm{E})^{\intercal}\bm{S}\rangle,
\label{eq:effective_S}
\end{equation}
which extends a result of Bergman \cite{Bergman:1983:SDL}.
The corresponding expression for the effective Hall tensor reads
\begin{equation}
\text{Cof}(\bm{\sigma}^{*})\bm{A}^{*}=\langle\text{Cof}(\bm{\sigma}\bm{E})^{\intercal}\bm{A}\rangle.
\label{eq:effective_A}
\end{equation}
A simplified version of this formula can be obtained by introducing the field $\bm{J}'=\bm{\sigma}\bm{E}(\bm{\sigma}^{*})^{-1}$, which links the macroscopic current density and the microscopic current density,
\begin{equation}
\bm{j}=\bm{J}'\langle\bm{j}\rangle,
\end{equation}
and in terms of which the effective Hall tensor is given by
\begin{equation}
\bm{A}^{*}=\langle\text{Cof}(\bm{J}')^{\intercal}\bm{A}\rangle.
\label{eq:effective_Ajp}
\end{equation}
The perturbative expressions (\ref{eq:effective_S}) and (\ref{eq:effective_Ajp}) will be used in the construction of the isotropic hierarchical laminate exhibiting a sign-inversion of the effective Hall coefficient and the discussion of the sign-inversion of individual components of the effective Hall tensor, respectively.

\subsection{Calculating the effective properties of laminates}
For the numerical calculations of the effective properties of the hierarchical laminates presented in this paper, the lamination formula obtained by Milton \cite{Milton:1990:CSP} and Zhikov \cite{Zhikov:1991:EHM} was used (note that an alternative lamination formula for arbitrary lamination directions has been derived by Murat and Tartar \cite{Tartar:1985:EFC}). For a detailed discussion of this formula, the reader is referred to Chap.\,9.3 of Ref.\onlinecite{Milton02}. Instead of working with the effective conductivity tensor, one introduces the tensors
\begin{align}
	\tilde{\bm{\mathscr{S}}}(\bm{b})&=\alpha\left(\alpha\bm{I}-\tilde{\bm{\sigma}}(\bm{b})\right)^{-1}\nonumber\\ \text{ and }\tilde{\bm{\mathscr{S}}}^*(\bm{b})&=\alpha\left(\alpha\bm{I}-\tilde{\bm{\sigma}}^*(\bm{b})\right)^{-1},
\end{align}
where $\alpha$ is a reference constant that can be freely chosen.
The formula, which has to be iteratively applied for each lamination step, reads
\begin{equation}
	\left(\tilde{\bm{\mathscr{S}}}^*(\bm{b})-\bm{\Gamma}_1(\bm{n})\right)^{-1}=\left\langle\left(\tilde{\mathscr{S}}(\bm{b})-\bm{\Gamma}_1(\bm{n})\right)^{-1}\right\rangle,
	\label{eq:lam_formula}
\end{equation}
where $\bm{\Gamma}_1(\bm{n})=\bm{n}\otimes\bm{n}$ with $\bm{n}$ being the direction of lamination, i.e., the direction perpendicular to the layers, and $\langle \cdot  \rangle$ denotes the volume average over the unit cell. Using this formula, the effective magnetic-field dependent conductivity tensor, $\tilde{\bm{\sigma}}(\bm{b})$, was numerically calculated for three mutually orthogonal directions of the magnetic field, which allows one to deduce all components of the effective $S$-tensor. The numerical calculations were carried out using the Python libraries NumPy \cite{Harris20} and SciPy \cite{Virta20}.

\subsection{Going beyond orthogonal laminates}

An orthogonal laminate is a hierarchical laminate whose directions of lamination are chosen from an orthogonal set of axes. Orthogonal laminates form an important class of laminates, particularly as they are known to attain many bounds on the effective properties of isotropic composites \cite{Briane04}---see, e.g., the recent work on the complex permittivity of three-dimensional two-phase composites in Ref.\,\onlinecite{Kern20}---and often also bounds on anisotropic effective properties---for example, as shown by Murat and Tartar \cite{Tartar:1985:EFC} and Lurie and Cherkaev \cite{Lurie86}, in the absence of a magnetic field, any (possibly anisotropic) effective conductivity than can be fundamentally attained by a three-dimensional composite that is made from two isotropic constituent materials and that has a fixed volume fraction is in fact attained by an orthogonal laminate. However, neither the effective Hall coefficient nor individual diagonal components of the effective Hall tensor of an orthogonal laminate can be sign-inverted if the constituent materials are isotropic.

This result follows immediately from Eq.\,(\ref{eq:effective_A}) and the fact that the matrix-valued electric field in an orthogonal laminate formed from isotropic constituent materials is positive diagonal if, w.l.o.g., the coordinate system is chosen such that its axes coincide with the lamination directions (see the proof of Thm.\,2.15 in Ref.\,\onlinecite{Briane04}). Hence, the Hall effect is exceptional in the sense that it requires one to consider composites beyond those that usually suffice. Specifically, in trying to find a laminate with a sign-inverted effective Hall coefficient (or a sign-inverted diagonal component of the effective Hall tensor), one is forced to allow at least one of the lamination directions to deviate from an orthogonal set of axes.

\section{Sign-Inversions of Components of the Effective Hall Tensor}

In this section, a laminate exhibiting a sign-inversion in one of the diagonal components of the effective Hall tensor is introduced. To see what such an inversion would correspond to in an experiment, we consider a long Hall bar made from a composite. We assume that the magnetic field is oriented along the $x_3$-direction, $\bm{b}=b_3\hat{\bm{x}}_3$, and that it is perpendicular to the Hall bar, in which the current is flowing parallel to the $x_1x_2$-plane, $\langle\bm{j}\rangle=(\langle j_1\rangle,\langle j_2\rangle,0)^{\intercal}$. Far away from the contacts, due to the deflection and subsequent accumulation of charge carriers, an electric field is built up that balances the magnetic part of the Lorentz force. The macroscopic counterpart of this so-called Hall electric field is given by
\begin{equation}
\langle\bm{e}_{\text{H}}\rangle=(\bm{A}^*\bm{b})\times\langle\bm{j}\rangle=b_3\begin{pmatrix}-A_{33}^*\langle j_2\rangle \\ A_{33}^*\langle j_1\rangle \\ A_{13}^*\langle j_2\rangle-A_{23}^*\langle j_1\rangle \end{pmatrix}.
\end{equation}
Hence, a sign-inversion in the component $A_{33}^*$ of the effective Hall tensor would translate to a sign-inversion in the $x_1$- and the $x_2$-component of the macroscopic Hall electric field, implying that, for this specific orientation of the composite, one would measure a sign-inverted Hall voltage.

\begin{figure}
	\includegraphics{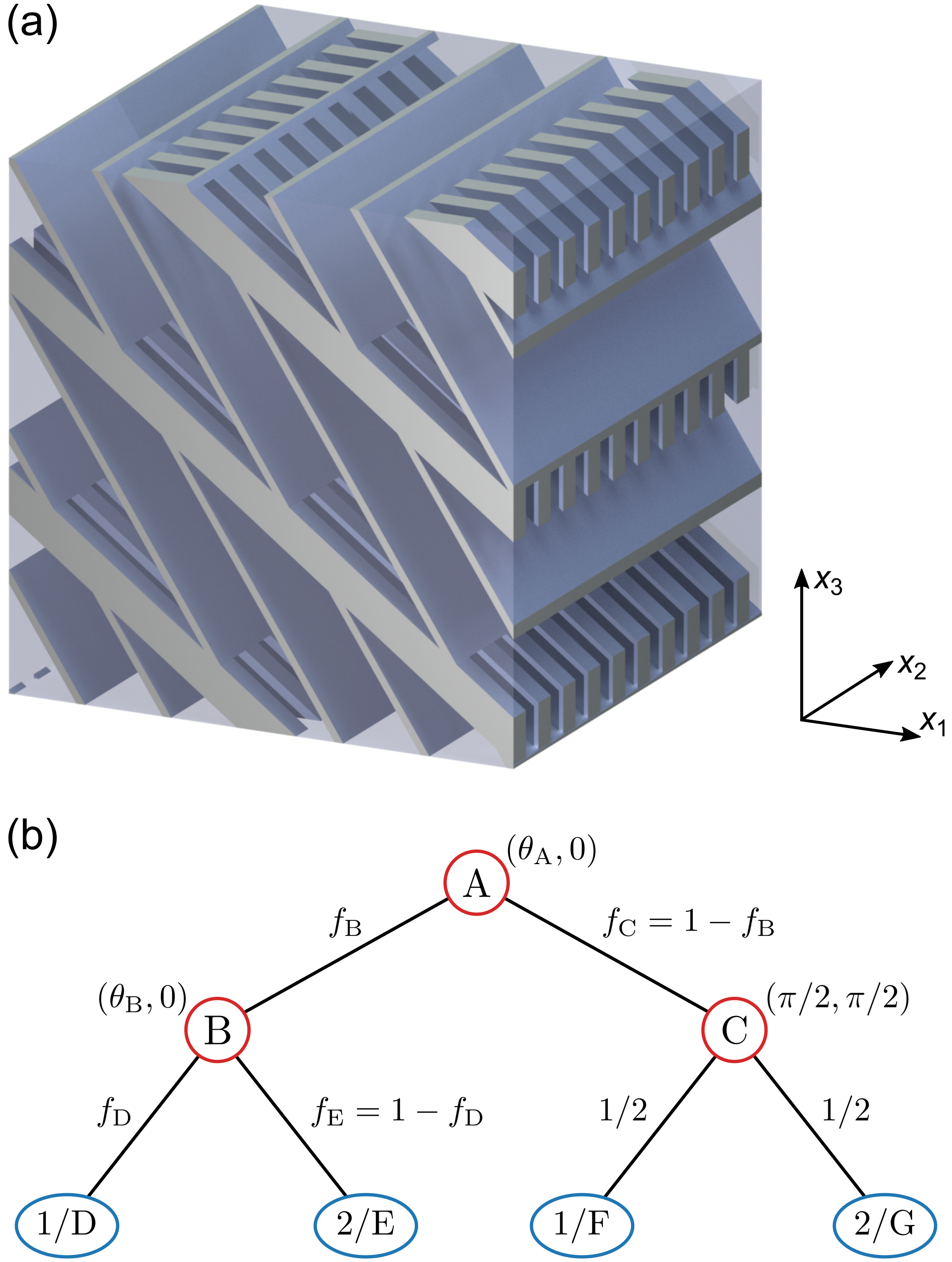}
	\caption{(a) Schematic illustration of the rank-two laminate showing a sign-inversion of the component $A_{33}^*$ of the effective Hall tensor. The laminate is made from two isotropic constituent materials. The first constituent material, which is weakly conducting, is depicted as semi-transparent. The second constituent material, which is highly conducting and has a non-zero Hall coefficient, is shown in grey. Unlike depicted here, the mathematical description assumes that the length scales of subsequent lamination steps are vastly separated. The values of the geometry parameters of the shown laminate are the result of the minimization of $A_{33}^*$ with the exception of the volume fraction $f_{\text{C}}$, which was enlarged for illustrative purposes. (b) Tree structure of the laminate. The nodes of the tree correspond to the different lamination steps (red circles) and the two constituent materials (blue ellipses). The volume fractions are assigned to the edges of the tree. The lamination directions are specified in spherical coordinates $(\theta,\phi)$, where $\theta$ and $\phi$ are the polar angle and the azimuthal angle, respectively, i.e., the convention common in physics is used.}
	\label{fig1}
\end{figure}
The laminate exhibiting a sign-inversion of the component $A^*_{33}$---or, depending on the orientation of the laminate any other diagonal component of $\bm{A}^*$---is a rank-two laminate made from two isotropic constituent materials only. The first constituent material has a low zero magnetic-field conductivity, $\sigma^{(1)}$, and zero Hall coefficient, $A^{(1)}=0$, while the second constituent material has a fairly large zero magnetic-field conductivity, $\sigma^{(2)}$, and a non-zero Hall coefficient, $A^{(2)}$. For the numerical calculations, it is assumed that $\sigma^{(1)}=1\,\si{\siemens\per\meter}$ and $\sigma^{(2)}=400\,\si{\siemens\per\meter}$. An illustration of the laminate and the corresponding tree structure are shown in Fig.\,\ref{fig1}. 

To obtain a structure exhibiting a strong sign-inversion, a minimization of the component $A_{33}^*$ of the effective Hall tensor was carried out. Four of the parameters of the laminate, the polar lamination angles $\theta_{\text{A}}$ and $\theta_{\text{B}}$ and the volume fractions $f_{\text{B}}$ and $f_{\text{D}}$, were allowed to vary. Using the SciPy \cite{Virta20} implementation of the downhill simplex method, the following parameter values were obtained
\begin{equation}
\theta_{\text{A}}=0.6390,~\theta_{\text{B}}=0.9924,~f_{\text{B}}=0.7898,~f_{\text{D}}=0.9915.
\end{equation}
The effective zmf\ conductivity tensor and the effective Hall tensor of the optimized laminate are given by
\begin{align}
&\bm{\sigma}^{\text{rank-two}}=\begin{pmatrix} 28.775 & 0 & -21.151 \\ 0 & 3.902 & 0 \\ -21.151 & 0 & 18.322 \end{pmatrix}\,\si{\siemens\per\meter}\\
&\text{and }\bm{A}^{\text{rank-two}}= \begin{pmatrix} 46.119 & 0 & 29.594 \\ 0 & 1.922 & 0 \\ -29.594 & 0 & -18.665 \end{pmatrix}A_{\text{H}}^{(2)},
\end{align}
respectively. Note that the effective Hall tensor shows the desired sign-inversion in the component $A_{33}^*$.

The reason underlying this sign-inversion can be seen in the fact that a macroscopic current flow along the $x_1$-direction leads to microscopic current flow in the phase 2/E that has a component in the opposite direction, i.e., in the fact that there is local inversion of current flow. Indeed, a numerical calculation gives $J'_{11}=-27.686<0$ in the phase 2/E. Note that, in order to arrive at this result, it is not necessary to go through all steps of the numerical calculation. It suffices to calculate the ``macroscopic'' current density in the sub-laminate B. Then, in order to determine the current flow in the phase 2/E, one can use the fact that the lamination formulae simplify vastly in the limit of large conductivity contrasts and volume fractions approaching zero or one (see the Appendix).

It should be pointed out that the local inversion of current flow is enabled by the oblique lamination angles. In each lamination step, the current density is successively rotated. A macroscopic current flow along the $x_1$-direction leads to a current flow roughly along the $x_3$-direction in the sub-laminate B and, ultimately, a current flow with a negative $x_1$-component in the phase 2/E.

This inverted current flow in the phase 2/E causes a local Hall voltage that is sign-inverted. As this phase extends throughout the laminate along the $x_2$-direction (and the effect is not compensated by the remainder of the laminate), the macroscopically measurable Hall voltage is sign-inverted too. 

One can understand the sign-inversion of $A_{33}^*$ also from Eq.\,(\ref{eq:effective_Ajp}). As the laminate is invariant with respect to arbitrary translations along the $x_2$-direction (except for the sublaminate B, which, however, has a diagonal effective conductivity tensor) and the constituent materials are isotropic, one has that $J'_{22}>0$ and $J'_{21}=J'_{12}=0$. Hence, the cofactor $\text{Cof}(\bm{J}')_{33}=J'_{11}J'_{22}-J'_{21}J'_{12}$, which determines the sign of the component $A_{33}^*$ of the effective Hall tensor of a composite made from isotropic constituent materials, is negative in the phase 2/E. As the value of this cofactor in the phase 2/G, in which it is not negative, is not large enough to compensate this effect, the effective Hall tensor component $A_{33}^*$ is sign-inverted. 

\section{Sign-Inversion of the Effective Hall Coefficient}
While the sign-inversion of individual components of the Hall tensor is already a curious effect, the main goal of this paper is to identify a laminate with isotropic properties that exhibits a sign-inversion of the effective Hall coefficient. This goal is reached in two steps: First, a laminate exhibiting a sign-inversion of the trace of the effective $S$-tensor is identified. Second, by extending an argument of Schulgasser's \cite{Schulgasser77}, it is shown that, if one has identified such a laminate, then one can use it to form an isotropic higher-rank laminate with a sign-inverted effective Hall coefficient.  
 
\subsection*{First step: Sign inversion of the trace of the\\ effective $S$-tensor}
To obtain a laminate with an effective $S$-tensor with a sign-inverted trace, an additional lamination step is used, i.e., a rank-three instead of a rank-two laminate is considered. This additional complexity seems to be necessary, as further numerical results suggest that, while the second invariant, i.e., the trace of the cofactor matrix, of the matrix-valued electric field, $\text{tr}(\text{Cof}(\bm{E}))$, in a rank-two laminate can locally be negative, which is a necessary condition for the desired sign-inversion (compare Eq.\,(\ref{eq:effective_S})), the desired sign-inversion itself cannot be obtained with two lamination steps only. 

A three-dimensional laminate with a locally negative value of $\text{tr}(\text{Cof}(\bm{E}))$ has been previously introduced by Briane and Nesi \cite{Briane04}. In the same paper, they had shown that the first invariant of the matrix-valued electric field, i.e. the determinant $\text{det}(\bm{E})$, is non-negative for any laminate irrespective of rank and dimension (see their Thm.\,2.13), which made it natural to ask whether this property extends to the second invariant. Their laminate and counterexample (illustrated in Fig.\,1 of their paper) has a geometry that is translationally invariant along one axis and is made from three constituent materials with well-separated conductivities (in Fig.\,1 of their paper four constituent materials are shown, but their number is reduced to three in Sec.\,6.2). 

\begin{figure*}
	\includegraphics{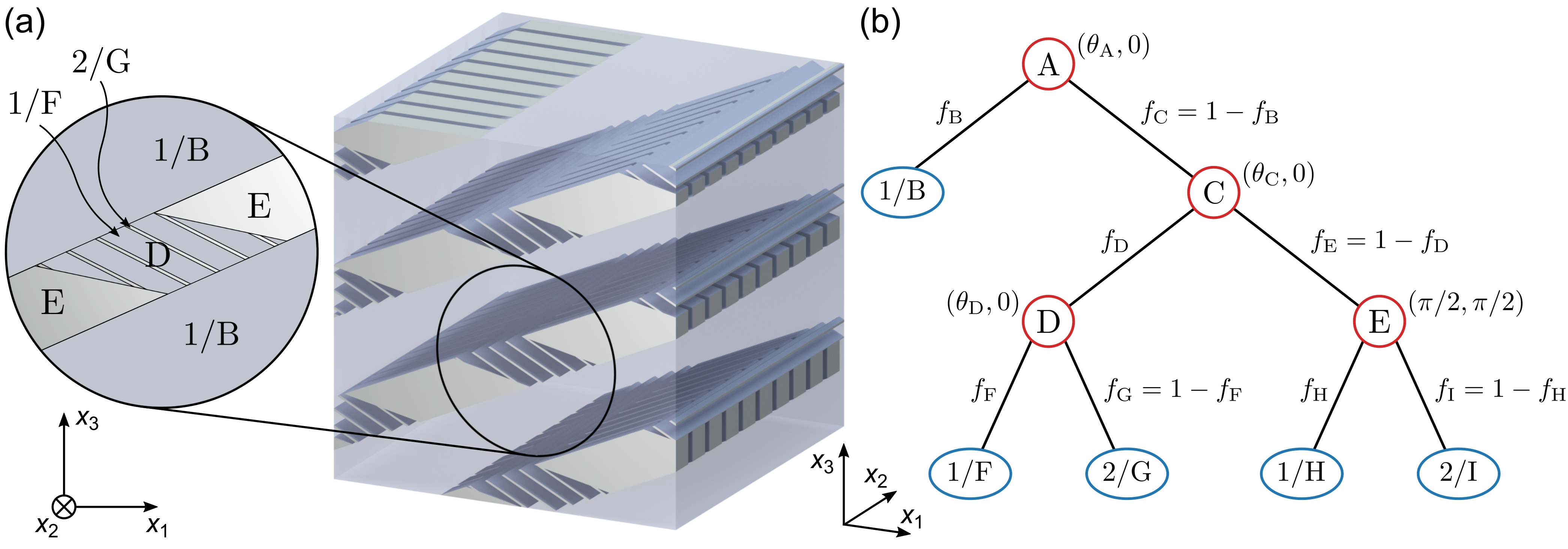}
	\caption{(a) Schematic illustration of the rank-three laminate exhibiting a sign-inversion of the trace of the effective $S$-tensor. The constituent materials are the same as those of the rank-two laminate. The detailed view on the left shows the structure of the rank-three laminate in the $x_1x_3$-plane. In the pictorial view on the right (but not in the detailed view on the left), the layering of the sub-laminate E perpendicular to the $x_2$-axis is visible. The parameter values of the depicted laminate are the result of the minimization of the component $A_{33}^*$ of the effective Hall tensor with the exception of the volume fraction $f_{\text{C}}$, which was enlarged for illustrative purposes. (b) Tree structure of the rank-three laminate including its parameters.}
	\label{fig2}
\end{figure*} 

The laminate introduced here, which is shown in Fig.\,2, is similar to theirs, but uses two constituent materials only. More precisely, the constituent materials are the same as in the last section. The laminate is formed from a rank-two laminate and the weakly conducting constituent material. It is translationally invariant along the $x_2$-direction with the exception of the sub-laminate E, which is formed by laminating the two constituent materials along the $x_2$-direction. This sub-laminate mimics an anisotropic homogeneous material with an intermediate conductivity in the $x_1x_3$-plane and a low conductivity along the $x_2$-direction. The remaining free geometry parameters are the polar lamination angles $\theta_{\text{A}}$, $\theta_{\text{C}}$, and $\theta_{\text{D}}$ and the volume fractions $f_{\text{B}}$, $f_{\text{D}}$, $f_{\text{F}}$, and $f_{\text{H}}$. 

As one would expect, there is no unique choice of these parameters leading to a sign-inversion of the trace of the effective $S$-tensor. To be able to compare the results with those obtained for the rank-two laminate in the last section, again the component $A_{33}^*$ of the effective Hall tensor is minimized. As it turns out, this choice not only results in an effective $S$-tensor with a sign-inverted trace, but also in a large modulus of the sign-inverted Hall coefficient of the isotropic laminate that will be formed from the rank-three laminate in the next section.  

Using the same minimization method and the same values for the constituent material parameters as in the last section, the following values for the geometry parameters were obtained
\begin{align}
\theta_{\text{A}}&=2.7224,~\theta_{\text{C}}=0.2078,~\theta_{\text{D}}=0.4712,\\f_{\text{B}}&=0.9590,~f_{\text{D}}=0.3834,~f_{\text{F}}=0.8670,~f_{\text{H}}=0.2652,\nonumber
\end{align}
and the corresponding effective conductivity tensor and effective Hall tensor are given by
\begin{align}
	&\bm{\sigma}^{\text{rank-three}}=\begin{pmatrix} 1.869 & 0 & 0.324 \\ 0 & 1.903 & 0 \\ 0.324 & 0 & 1.166 \end{pmatrix}\,\si{\siemens\per\meter} \\&\text{and }\bm{A}^{\text{rank-three}}=\begin{pmatrix} 15.010 & 0 & 27.861 \\ 0 & 0.699 & 0 \\ -27.861 & 0 & -51.444 \end{pmatrix}A_{\text{H}}^{(2)},
\end{align}
respectively. Notably, the inversion is much stronger than in the rank-two case in the sense that the minimized sign-inverted component of the effective Hall tensor has a much larger modulus. The corresponding effective $S$-tensor,
\begin{align}
\bm{S}^{\text{rank-three}}=\begin{pmatrix} 3.155 & 0 & 5.845 \\ 0 & 0.091 & 0 \\ -6.770 & 0 & -12.507 \end{pmatrix}S^{(2)}\cdot 10^{-4},
\end{align}
clearly has a sign-inverted trace implying that the first step of the construction is completed. Note that, analogously to the discussion of the rank-two laminate, the sign-inversion of $S_{33}^*$ can be attributed to a sign-inversion of the component $E_{11}$ of the matrix-valued electric field in the phase 2/G.

\subsection*{Second step: Obtaining an isotropic laminate}

It will now be shown that from any anisotropic material, i.e., any material with an anisotropic (effective) zmf\ conductivity tensor, $\tilde{\bm{\sigma}}^{\text{aniso}}$, and/or anisotropic (effective) $S$-tensor, $\bm{S}^{\text{aniso}}$, one can form an isotropic laminate with effective zmf\ conductivity $\tilde{\sigma}^{\text{iso}}=\text{tr}(\tilde{\bm{\sigma}}^{\text{aniso}})/3$ and effective $S$-coefficient $S^{\text{iso}}=\text{tr}(\bm{S}^{\text{aniso}})/3$. As we will see, this result implies that one can form an isotropic laminate with a sign-inverted effective Hall coefficient from the rank-three laminate obtained in the previous section.\\

The construction is based on a so-called Schulgasser laminate \cite{Schulgasser77}, see also Sec.\,22.2 in Ref.\,\onlinecite{Milton02}. Schulgasser proposed a specific way of forming a laminate that is isotropic (as long as only the conductivity is concerned) from an anisotropic material. A Schulgasser laminate has the specific property that its effective conductivity is the largest among all isotropic polycrystals that can possibly be formed from the anisotropic material. The key idea is to repeatedly permute diagonal components (via a rotation) and subsequently take an arithmetic mean (via a lamination). In the following description, the coordinate system is chosen such that the zmf\ conductivity tensor of the anisotropic material is diagonal. 

Schulgasser's original construction involves three lamination steps on widely separated length scales:

In the first step, the anisotropic material is laminated with a rotated copy of itself. The axis of rotation can be chosen arbitrarily from one of the three coordinate axes and the angle of rotation is $90^\circ$. Consequently, the zmf\ conductivity tensor of the rotated copy is still diagonal. The effect of the rotation is simply a transposition of the two diagonal components perpendicular to the axis of rotation while the component parallel to the axis of rotation is unaffected. The direction of lamination is identical to the axis of rotation. As a result of the fact that the zmf\ conductivity tensors of the anisotropic material and its rotated copy are diagonal and their components in the direction of lamination are identical, the general expression for the effective zmf\ conductivity tensor of the resulting laminate reduces to a (volume-fraction weighted) arithmetic mean. 

In the two subsequent steps, the procedure is analogous, but it is now the laminate obtained in the previous step that is laminated with a rotated version of itself. Again, the rotation and lamination correspond to a transposition of two of the diagonal components and taking an arithmetic mean, respectively. If the volume fractions and the axes of rotation are chosen suitably, the zmf\ conductivity tensor of the laminate obtained in the last step, i.e., the Schulgasser laminate, is given by
\begin{equation}
\tilde{\bm{\sigma}}^{\text{Schulg}}=\text{tr}(\tilde{\bm{\sigma}}^{\text{aniso}})/3.
\label{eq:conductivity_Schulgasser}
\end{equation}

We can now derive an expression for the effective $S$-tensor of a Schulgasser laminate. First note that the $S$-tensor is a (proper) rank-two tensor, see Ref.\,\onlinecite{LandauCont}, i.e., it has the same transformational properties as the zmf\ conductivity tensor, in particular under rotation. However, the $S$-tensor and the zmf\ conductivity tensor of the anisotropic material may not be simultaneously diagonalizable, i.e., the $S$-tensor may have non-zero non-diagonal components in the chosen coordinate system. Next, we again consider the general perturbative expression, Eq.\,(\ref{eq:effective_S}). As in each lamination step, the effective zmf\ conductivity tensors of the laminate obtained in the previous step (in the first step, the zmf\ conductivity tensor of the anisotropic material) and of the corresponding rotated copy are diagonal and their components in the direction of lamination are equal, the microscopic electric field is given by the identity (see the Appendix). Hence, the effective $S$-tensor of the laminate obtained in each step is given by the arithmetic mean of the laminate obtained in the previous step and its rotated copy and, ultimately, the expression for the diagonal components of the effective $S$-tensor of the Schulgasser laminate is analogous to the expression for the effective zmf\ conductivity tensor, Eq.\,(\ref{eq:conductivity_Schulgasser}),
\begin{equation}
S^{\text{Schulg}}_{ii} = \text{tr}(\bm{S}^{\text{aniso}})/3 \text{ for every } i \in \lbrace 1,2,3\rbrace.
\end{equation}

Potentially non-zero non-diagonal components of the $S$-tensor of the Schulgasser laminate can be eliminated via two additional lamination steps using the fact that if the microscopic zmf\ conductivity is constant, then the expression for the effective $S$-tensor reduces to an arithmetic mean. In the first additional lamination step, the Schulgasser laminate is laminated in equal proportions with a copy of itself that is reflected at the $x_2x_3$-plane. In the second additional lamination step, it is the laminate obtained in the first step that is laminated in equal proportions with a copy of itself, the plane of reflection being the $x_1x_3$-plane. The resulting rank-five laminate is isotropic and has the desired effective conductivity and effective Hall coefficient.

Finally, this construction is used to obtain an isotropic hierarchical laminate with a sign-inverted effective Hall coefficient. Assume that the anisotropic material is the rank-three laminate obtained in the last section. Then, the effective Hall coefficient of the isotropic laminate is given by
\begin{equation}
A^{\text{iso}} = -\frac{S^{\text{iso}}}{(\tilde{\sigma}^{\text{iso}})^2} = -\frac{\text{tr}(\bm{S}^{\text{rank-three}})}{3(\tilde{\sigma}^{\text{iso}})^2}.
\end{equation}
As the trace of the $S$-tensor of the rank-three laminate, $\text{tr}(\bm{S}^{\text{rank-three}})$, is sign-inverted, the Hall coefficient of the isotropic laminate, $A^{\text{iso}}$, is sign-inverted as well. Using the parameters obtained for the rank-three laminate in the last section, the effective conductivity and the effective Hall coefficient of the isotropic hierarchical laminate are given by
\begin{equation}
\tilde{\sigma}^{\text{iso}}= 1.646\,\si{\siemens\per\meter} \text{ and }A^{\text{iso}} = -18.230\,A^{(2)},
\end{equation} 
respectively.

A slightly smaller value of the effective Hall coefficient of the isotropic hierarchical laminate, $A^{\text{iso}}$, can be obtained by minimizing it directly instead of minimizing one of the components of the effective Hall tensor of the rank-three laminate. During the minimization, one of the polar lamination angles of the rank-three laminate can be held fixed as adding the same offset to $\theta_{\text{A}}$, $\theta_{\text{C}}$, and $\theta_{\text{D}}$ corresponds to a rotation of the rank-three laminate in the $x_1x_3$-plane, which does not affect the effective properties of the isotropic laminate. The resulting effective conductivity and effective Hall coefficient of the isotropic hierarchical laminate are given by
\begin{equation}
	\tilde{\sigma}^{\text{iso}}= 1.987\,\si{\siemens\per\meter} \text{ and }A^{\text{iso}} = -20.826 A^{(2)},
\end{equation} 
respectively.

\subsection*{Evaluating the performance of the laminates}

Having established that there are isotropic laminates exhibiting a sign-inversion of the effective Hall coefficient, it is now investigated how well the identified laminates perform. The effective Hall coefficient is used as the figure of merit. Typically, a Hall coefficient with a large modulus is desirable as a Hall bar made from such a material will have a large current-related sensitivity \cite{Popovic03}\footnote{However, it should be noted that while composites can exhibit an enhanced effective Hall coefficient, this enhancement can be readily mimicked by making the Hall bar thinner.}.

A natural way of assessing the performance of a composite is to compare its effective properties with a bound. For the structures considered here, a suitable bound has been derived by Briane and Milton \cite{Briane09Giant}. They have shown that each component of the effective Hall tensor is bounded as follows, see Eq.\,(2.30) in their paper,
\begin{equation}
\left|A_{ij}^*\right|\leq 6 \frac{\beta_{\text{H}}}{\alpha}a_{\text{H}}\text{ for any }i,j\in\lbrace 1,2,3\rbrace,
\label{eq:bound_giant}
\end{equation}
where $\alpha$ is a lower bound on the conductivity of the constituent materials, $\beta_{\text{H}}$ is an upper bound on the conductivity of the constituent materials with non-zero Hall coefficient, and $a_{\text{H}}$ is an upper bound on the Hall coefficient of the constituent materials. 

From a glance at the results obtained in the previous sections, it becomes clear that the isotropic hierarchical laminate as well as the anisotropic rank-three laminate are far from attaining this bound. Consequently, it might be possible to find yet better composites (which may or may not be laminates) and/or to derive an improved bound. An indication that there should be a tighter bound is the fact that the bound in question does not differentiate between the sign-inverted and the non-sign-inverted regime. Intuitively, one would expect that a sign-inverted effective Hall coefficient (or component of the effective Hall tensor) is more strongly constrained. Furthermore, one would certainly expect that there is a tighter bound for isotropic composites, as the bound\,(\ref{eq:bound_giant}) was derived for individual components of the Hall tensor implying that isotropy was not an assumption entering its derivation. 

\begin{figure}[h!]
	\includegraphics{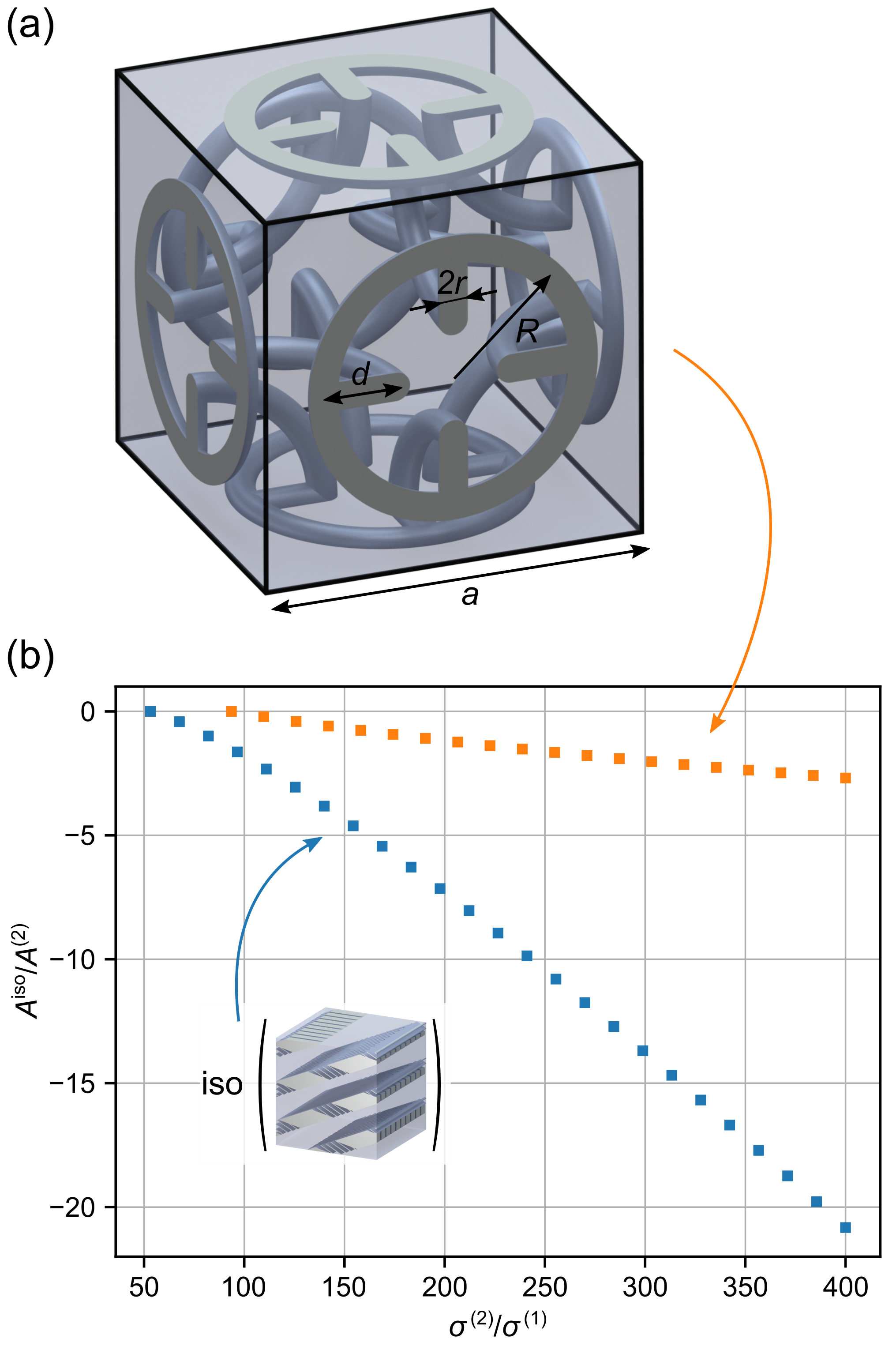}
	\caption{Evaluating the performance of the isotropic hierarchical laminate by comparing it with a modified version of the chain-mail-inspired microstructure. (a) Illustration of a unit cell of the modified version of the chain-mail-inspired microstructure. As opposed to the microstructure studied in Ref.\,\onlinecite{Kadic:2015:HES}, the periodic arrangement tori and cylinders is embedded in a weakly conducting surrounding medium. The lattice constant, $a$ is determined by the major radius of the tori, $R$, and the distance parameter, $d$, via $a=4R+2d$ (using the convention that $d$ is negative if the tori are intertwined). (b) Numerical results for the effective Hall coefficient of the optimized structures as a function of the ratio of conductivities of the two constituent materials. The blue squares correspond to the isotropic hierarchical laminate, as indicated by the inset. The orange squares correspond to the chain-mail-inspired composite. The smallest conductivity ratio for which a sign-inversion is obtained is $54$ and $94$ for the isotropic hierarchical laminate and the chain-mail-inspired composite, respectively.}
	\label{fig3}
\end{figure} 
 
Another way of assessing the performance of a given composite geometry is to compare it with known microstructures. In the following, the isotropic hierarchical laminate is compared with the simplified version of the chain-mail-inspired microstructure that was introduced in Ref.\,\onlinecite{Kadic:2015:HES}. This microstructure consists of a periodic arrangement of intertwined solid tori that are connected by short cylinders. The geometry parameters of the structure are the major radius of the tori, $R$, the minor radius of the tori, $r$, which is simultaneously the radius of the cylinders, and the signed distance between the axial circles of neighboring tori, $d$. In the paper introducing this simplified version \cite{Kadic:2015:HES}, as well as in the later experimental realizations \cite{Kern:2016:EES, Kern19}, the structure was made from a single constituent material. Here, while the tori and cylinders are still made from an isotropic material with non-zero Hall coefficient, the microstructure is embedded in a weakly conducting isotropic material with zero Hall coefficient. More precisely, for the comparison the same two constituent materials are used for both the laminate and the chain-mail-inspired composite. A corresponding unit cell is illustrated in Fig.\,\ref{fig3}(a).

The quantity that is compared is the smallest attainable value of the effective Hall coefficient of the two structures as a function of the ratio of conductivities. For the isotropic hierarchical laminate, the minimization of $A^{\text{iso}}$ already described in the last section was repeated for each value of $\sigma^{(2)}/\sigma^{(1)}$. For the chain-mail-inspired composite, the effective properties were deduced from finite element calculations that were carried out as described in Ref.\,\onlinecite{Kern18-1}. In the latter case, the effective Hall coefficient was minimized as a function of the geometry parameters $d/R$ and $r/R$. Varying only these two parameters is sufficient, as scaling the parameter $R$ while keeping $d/R$ and $r/R$ fixed corresponds to scaling the microstructure as a whole, which does not affect its effective properties. 

The numerically calculated values of the relative effective Hall coefficient of the optimized structures as a function of $\sigma^{(2)}/\sigma^{(1)}$ are shown in Fig.\,\ref{fig3}(b). For both structures, the general behavior is the same. A sign-inverted effective Hall coefficient is obtained only if the ratio of conductivities exceeds a certain threshold. As the ratio of conductivities increases further, the modulus of the effective Hall coefficient of the optimized structures increases as well. This behavior is typical for composites in the sense that obtaining a wide range of attainable effective properties and especially obtaining exotic effective properties generally requires a sufficiently large contrast in the properties of the constituent materials. However, both in regard to the value of the threshold above which a sign-inversion is obtained as well as in regard to the value of the obtained effective Hall coefficient at larger conductivity ratios, the laminate significantly outperforms the chain-mail-inspired composite. 

\section{Conclusions}
The sign-inversion of the effective Hall coefficient in composites was studied. In contrast to the seminal mathematical work \cite{Briane:2004:CSC, Briane09} and the later theoretical \cite{Kadic:2015:HES, Kern18-1} and experimental \cite{Kern:2016:EES, Kern19} studies that built upon it, the present paper was concerned with the question whether this effect can occur in hierarchical laminates.

First, a rank-two laminate exhibiting a sign-inversion of one of the diagonal components of its effective Hall tensor was introduced. If such a composite is used to construct a conventional Hall bar and the orientation of the composite is suitably chosen, the measured Hall voltage is sign-inverted, feigning that its charge carriers are oppositely charged.

Second, an example of an isotropic hierarchical laminate with a sign-inverted effective Hall coefficient was given. Its construction was based on a rank-three laminate exhibiting a sign-inversion of the trace of its effective $S$-tensor. Using an idea of Schulgasser's, additional lamination steps were used to obtain isotropic behavior. In this case, the Hall voltage measured on a conventional Hall bar is sign-inverted independently of the orientation of the composite.

In comparison with the inception of the previously introduced chain-mail-inspired structure, the isotropic hierarchical laminate was obtained with relative conceptional and computational ease. Furthermore, the sign-inversion was obtained for a smaller ratio of the conductivities of the constituent materials and the optimization resulted in a larger modulus of the sign-inverted effective Hall coefficient.

These results show that the effective properties of hierarchical laminates extend yet farther than expected, indicating that their potential has not been fully exploited yet. As, additionally, their effective behavior can be readily calculated, hierarchical laminates constitute a useful tool for establishing that certain effective properties can be obtained in principle (see also Ref.\,\onlinecite{Milton86}). If one aims to find microstructures that can be fabricated more easily, the identified hierarchical laminates can then serve as a benchmark.

\begin{acknowledgments}
This work was initiated and partly carried out while the author held postdoctoral positions at the University of Utah and the Advanced Science Research Center, CUNY, respectively. The author is very grateful to Graeme Milton (University of Utah) for many helpful and stimulating discussions and for support during a stay in Salt Lake City. Helpful remarks by Andrea Al\`{u} (Advanced Science Research Center, CUNY), Muamer Kadic (FEMTO-ST, Universit\'{e} Bourgogne Franche-Comt\'{e}), Aaron Welters (Florida Institute of Technology), Vincenzo Nesi (Sapienza University of Rome), and Marc Briane (INSA de Rennes) are kindly acknowledged.
	
The author thanks the National Science Foundation for partial support through Research Grants DMS-1814854 and DMS-2107926.
\end{acknowledgments}

\appendix*
\section{Expressions for the microscopic fields}
We briefly discuss the expressions for the microscopic fields in a laminate formed from two phases with zmf\ conductivity tensors $\bm{\sigma}^{(\alpha)}$ and $\bm{\sigma}^{(\beta)}$. It is assumed that there is no magnetic field present. The coordinate system is chosen such that its $x_1$-axis aligns with the direction of lamination (i.e., the direction perpendicular to the layers). 
\begin{figure}[h]
	\includegraphics{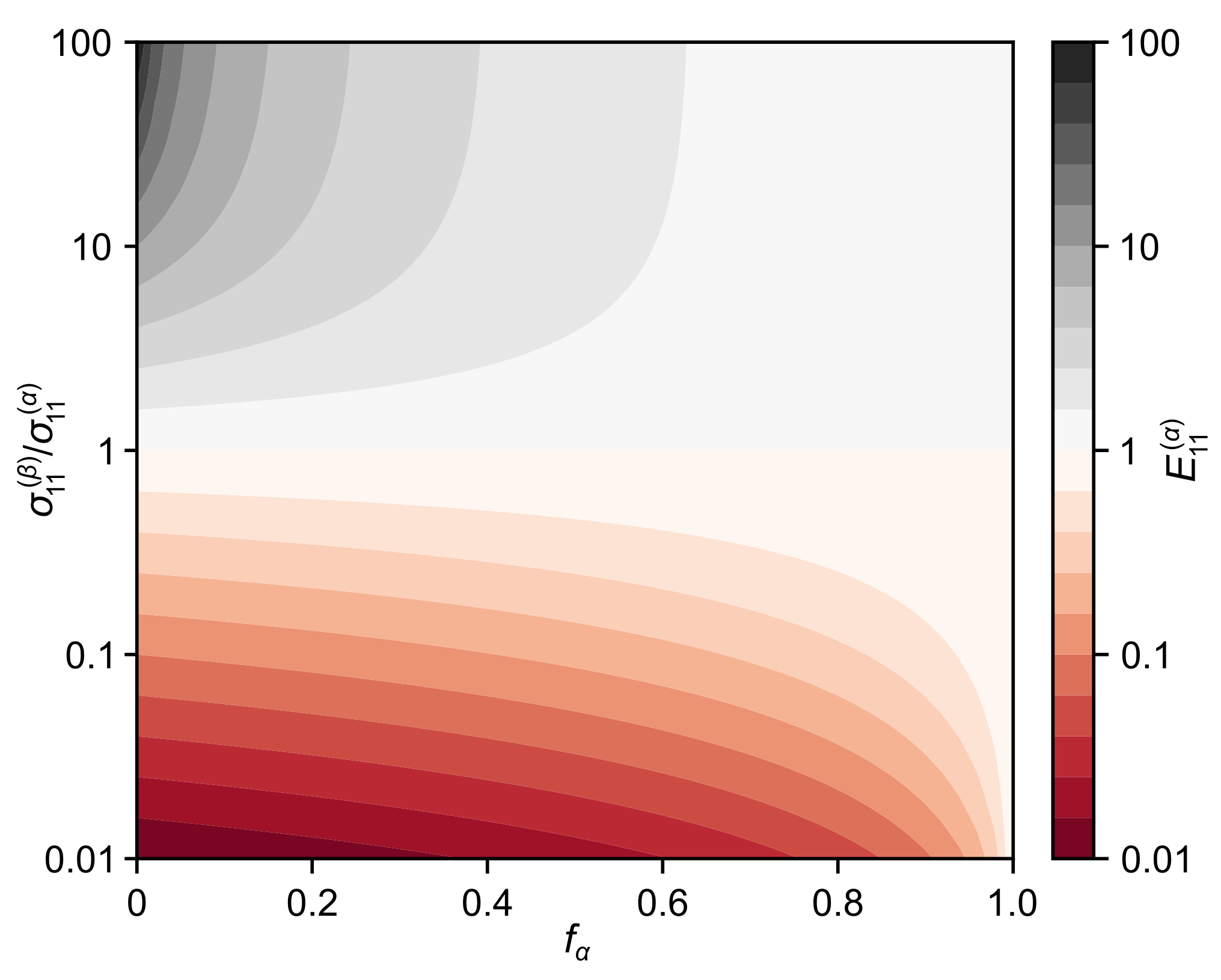}
	\caption{Dependence of the component $E_{11}^{(\alpha)}$ of the matrix-valued electric field in one of the phases of a two-phase laminate on the volume fraction, $f_{\alpha}$, and the ratio of conductivities, $\sigma_{11}^{(\beta)}/\sigma_{11}^{(\alpha)}$. In the three regimes in which $E_{11}^{(\alpha)}$ takes very large or very small values or values close to one, the link between the macroscopic and the microscopic electric field is particularly simple. If one relabels the vertical axis as $\sigma_{22}^{(\beta)}/\sigma_{22}^{(\alpha)}$ or $\sigma_{33}^{(\beta)}/\sigma_{33}^{(\alpha)}$, under the assumption that the conductivity tensors of the phases are diagonal, the plot shows the behavior of $J_{22}'^{(\alpha)}$ or $J_{33}'^{(\alpha)}$, respectively, which link a given macroscopic current density to the corresponding microscopic current density in phase $\alpha$.}
	\label{fig4}
\end{figure} 
Then, the fields satisfy the following set of equations, see Chap.\,9 in Ref.\,\onlinecite{Milton02},
\begin{align}
&f_{\alpha}\bm{e}^{(\alpha)}+f_{\beta}\bm{e}^{(\beta)}=\langle\bm{e}\rangle \text{ with }f_{\beta}=1-f_{\alpha},\nonumber\\
&j_1^{(\alpha)}=j_1^{(\beta)},\text{ and }e_k^{(\alpha)}=e_k^{(\beta)}=\langle e_k\rangle \text{ for }k= 2,\,3.
\end{align}
One can solve these equations to obtain the microscopic electric field in phase $\alpha$ and phase $\beta$, $\bm{e}^{(\alpha)}$ and $\bm{e}^{(\beta)}$, respectively. 
For the diagonal components of the matrix-valued electric field in phase $\alpha$, one finds
\begin{equation}
E_{11}^{(\alpha)}=\left(f_{\alpha}+f_{\beta}\frac{\sigma_{11}^{(\alpha)}}{\sigma_{11}^{(\beta)}}\right)^{-1}\\
\text{and }E_{22}^{(\alpha)}=E_{33}^{(\alpha)}=1.
\label{eq:append_matr_el}
\end{equation}
Under the additional assumption that the zmf\ conductivity tensors of the two phases are diagonal, these are the only non-vanishing components. The mathematical treatment is considerably simplified in the following three regimes:
\begin{itemize}
\item For $f_{\alpha}\rightarrow 0,~\sigma_{11}^{(\beta)}/\sigma_{11}^{(\alpha)}\rightarrow 0$, one has $E_{11}^{(\alpha)}\rightarrow 0$, i.e., the macroscopic electric field is projected onto the plane perpendicular to the direction of lamination.
\item For $f_{\alpha}\rightarrow 0,~\sigma_{11}^{(\beta)}/\sigma_{11}^{(\alpha)}\rightarrow \infty$, one has $E_{11}^{(\alpha)}\rightarrow \infty$, i.e., the $x_1$-component of the microscopic electric field is strongly enhanced relative to the same component of the macroscopic electric field. If the latter is non-zero, then the microscopic electric field becomes almost parallel to the direction of lamination.
\item For $f_{\beta}\rightarrow 0,~f_{\beta}\,\sigma_{11}^{(\alpha)}/\sigma_{11}^{(\beta)}\rightarrow 0$ or $\sigma_{11}^{(\beta)}/\sigma_{11}^{(\alpha)}\rightarrow 1$, one has $E_{11}^{(\alpha)}\rightarrow 1$, i.e., the microscopic electric field approaches the macroscopic electric field.
\end{itemize}
A plot of the component $E_{11}^{(\alpha)}$ as a function of $f_{\alpha}$ and $\sigma_{11}^{(\beta)}/\sigma_{11}^{(\alpha)}$ in which these three regimes are readily apparent is shown in Fig.\,\ref{fig4}. The obtained simple links between the direction of lamination and the microscopic electric field in these regimes can be used to intuitively design hierarchical laminates in which the microscopic electric field shows a certain desired behavior.

For the electric current density, a similar expression can be given,
\begin{equation}
\bm{J}'^{(\alpha)}=\text{diag}\left(1,\left(f_{\alpha}+f_{\beta}\frac{\sigma_{22}^{(\beta)}}{\sigma_{22}^{(\alpha)}}\right)^{-1},\left(f_{\alpha}+f_{\beta}\frac{\sigma_{33}^{(\beta)}}{\sigma_{33}^{(\alpha)}}\right)^{-1}\right),
\end{equation}
where it was again assumed that the zmf\ conductivity tensors of the phases are diagonal. The discussion of the limiting behavior is then analogous to the one for the electric field.  
%

\end{document}